\documentclass{moriond}
\usepackage{amsmath,amssymb}

\def\Journal#1#2#3#4{{#1} {\bf #2}, #3 (#4)}

\def\NPB{{\em Nucl. Phys.} B}

\def\PRL{\em Phys. Rev. Lett.}

\def\be{\begin{equation}}
\def\ee{\end{equation}}
\def\bea{\begin{eqnarray}}
\def\eea{\end{eqnarray}}

\newcommand{\ep}{\epsilon}
\newcommand{\primed}{^{\prime}}
\def\beq#1\eeq{\begin{align}#1\end{align}}

\newcommand{\tev}{\,\mbox{TeV}}

\renewcommand{\Re}{\textrm{Re}\,}

\newcommand{\Photo}{}
\begin{document}
\hfill TTP17--025\\
\vspace*{1.2cm}

\title{ \boldmath{CORRELATIONS OF $\epsilon^{\prime}_K/\epsilon_K$ WITH $K \to \pi \nu \overline{\nu}$} IN MODELS OF NEW PHYSICS 
\footnote[0]{Presented at the 52th Rencontres de Moriond electroweak interactions and unified theories, La Thuile, Italy, 18-25 March, 2017.}
}

\author{ T.~KITAHARA }

\address{Institute for Theoretical Particle Physics (TTP), 
Karlsruhe Institute of Technology, Wolfgang-Gaede-Stra\ss e 1, 
 76128 Karlsruhe, Germany\\
Institute for Nuclear Physics (IKP), Karlsruhe Institute of
  Technology,\\ Hermann-von-Helmholtz-Platz 1, 76344
  Eggenstein-Leopoldshafen, Germany}

\maketitle\abstracts{
Recent calculations have pointed to a 2.8\,$\sigma$ tension between data on $\epsilon^{\prime}_K / \epsilon_K$ and the standard-model (SM) prediction. 
Several new physics (NP) models can explain this discrepancy,
and such NP models are likely to predict deviations of $\mathcal{B}(K\to \pi \nu \overline{\nu})$ from the SM predictions,  which can be probed precisely in the near future 
by NA62 and KOTO experiments.
We present correlations between $\epsilon^{\prime}_K / \epsilon_K$ and $\mathcal{B}(K\to \pi \nu \overline{\nu})$ 
in two types of NP scenarios: a box dominated scenario and a $Z$-penguin dominated one.
 It is shown that  different correlations are predicted and the future precision measurements  of 
  $K \to \pi \nu \overline{\nu}$ can distinguish both scenarios.
}

\section{Introduction}

CP violating flavor-changing neutral current decays of $K$ mesons are extremely sensitive to new physics (NP) and can probe virtual effects of particles with masses far above the reach of the Large Hadron  Collider.
 Prime examples of such observables are
$\ep_K\primed$ measuring direct CP violation in $K\to
\pi\pi$ decays and  $\mathcal{B}(K_L\to\pi^{0}\nu\overline{\nu})$. 
Until recently, large theoretical uncertainties precluded reliable
predictions for $ \epsilon _K^{\prime}$. 
Although standard-model (SM) predictions of $ \epsilon _K^{\prime}$ using chiral  perturbation theory  are consistent with the experimental value, their theoretical uncertainties are large.
In contrast, calculation by the dual QCD approach \cite{dual}
 finds the SM value much below the experimental one.
  A major breakthrough has been the recent lattice-QCD calculation of the hadronic matrix elements by RBC-UKQCD collaboration \cite{Bai:2015nea}, which gives support to 
the latter  result. 
The SM value at the next-to-leading order  divided by the indirect CP violating measure $\epsilon_K$ is~\cite{Kitahara:2016nld}
\begin{equation}
\Re  (\epsilon_K\primed / \epsilon_K )_{\rm SM} = 
(1.06 \pm 4.66_{\rm Lattice} \pm 1.91_{\rm NNLO} \pm 0.59_{\rm IV} \pm 0.23_{m_t} ) \times 10^{-4},
\label{eq:sm}
\end{equation}
which is consistent with
  $(\epsilon_K\primed/\epsilon_K)_{\rm SM} =
 (1.9 \pm 4.5) \times 10^{-4}$ given by Buras {\it et al}\,~\cite{Buras:2015yba}.\,\footnote{Other estimations of the SM value  are listed in Kitahara  {\it et al}\,~\cite{Kitahara:2016ftn}.}
Both results are based on the lattice numbers, and further use CP-conserving $K\to
\pi\pi $ data to constrain some of the hadronic matrix elements involved. 
Compared to the world average of the experimental results~\cite{Olive:2016xmw},
\begin{equation} 
\Re(\epsilon_K\primed/\epsilon_K)_{\rm exp} =
 (16.6 \pm 2.3) \times 10^{-4},
   \end{equation} 
the SM prediction lies below the experimental value by 2.8\,$\sigma$.

Several NP models including supersymmetry (SUSY) can explain this discrepancy. It is known that such NP models are likely to predict deviations of the kaon rare decay branching ratios from the SM predictions, especially $\mathcal{B}(K\to \pi \nu \overline{\nu})$ which can be probed precisely in the near future by NA62 and KOTO experiments.\footnote{
The correlations between $\epsilon^{\prime}_K / \epsilon_K$, $\mathcal{B}(K\to \pi \nu \overline{\nu})$ and $\epsilon_K$ through the CKM components in the SM are discussed in Ref.~\cite{Lehner:2015jga}.}
In this contribution, we present 
correlations between $\epsilon^{\prime}_K / \epsilon_K$ and $\mathcal{B}(K\to \pi \nu \overline{\nu})$ 
in two types of NP scenarios: a box dominated scenario and a $Z$-penguin dominated one.

\section{Box dominated (Trojan penguin) scenario}

%%%%%%%%%%%%%%%%%%%%%%%%%%%%%%%%%%%%%%%%%%%%
\begin{figure}
\centerline{\includegraphics[width=0.4\textwidth, bb= 0 0 356 350]{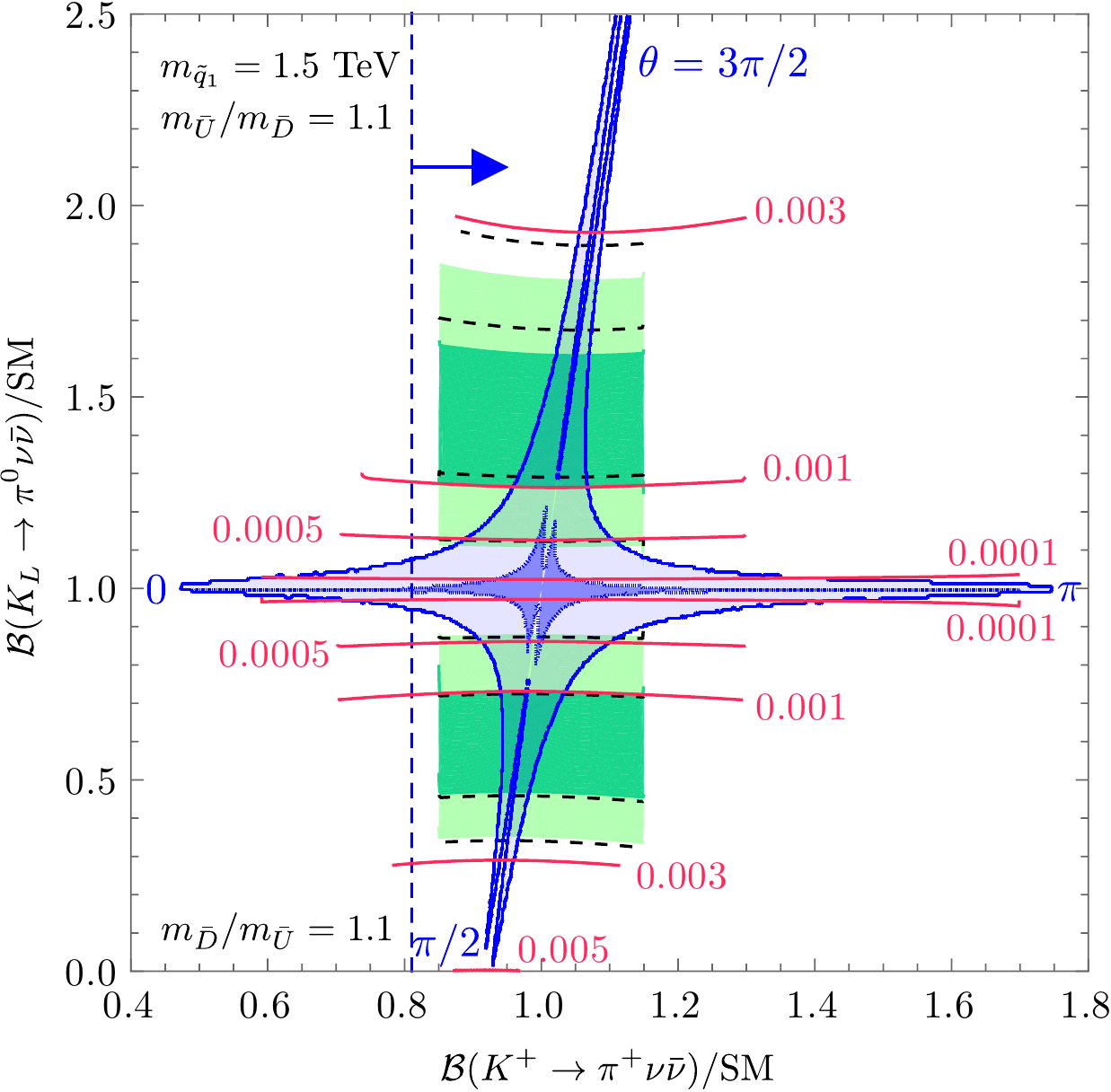}~~~~
\includegraphics[width=0.4\textwidth, bb= 0 0 352 345]{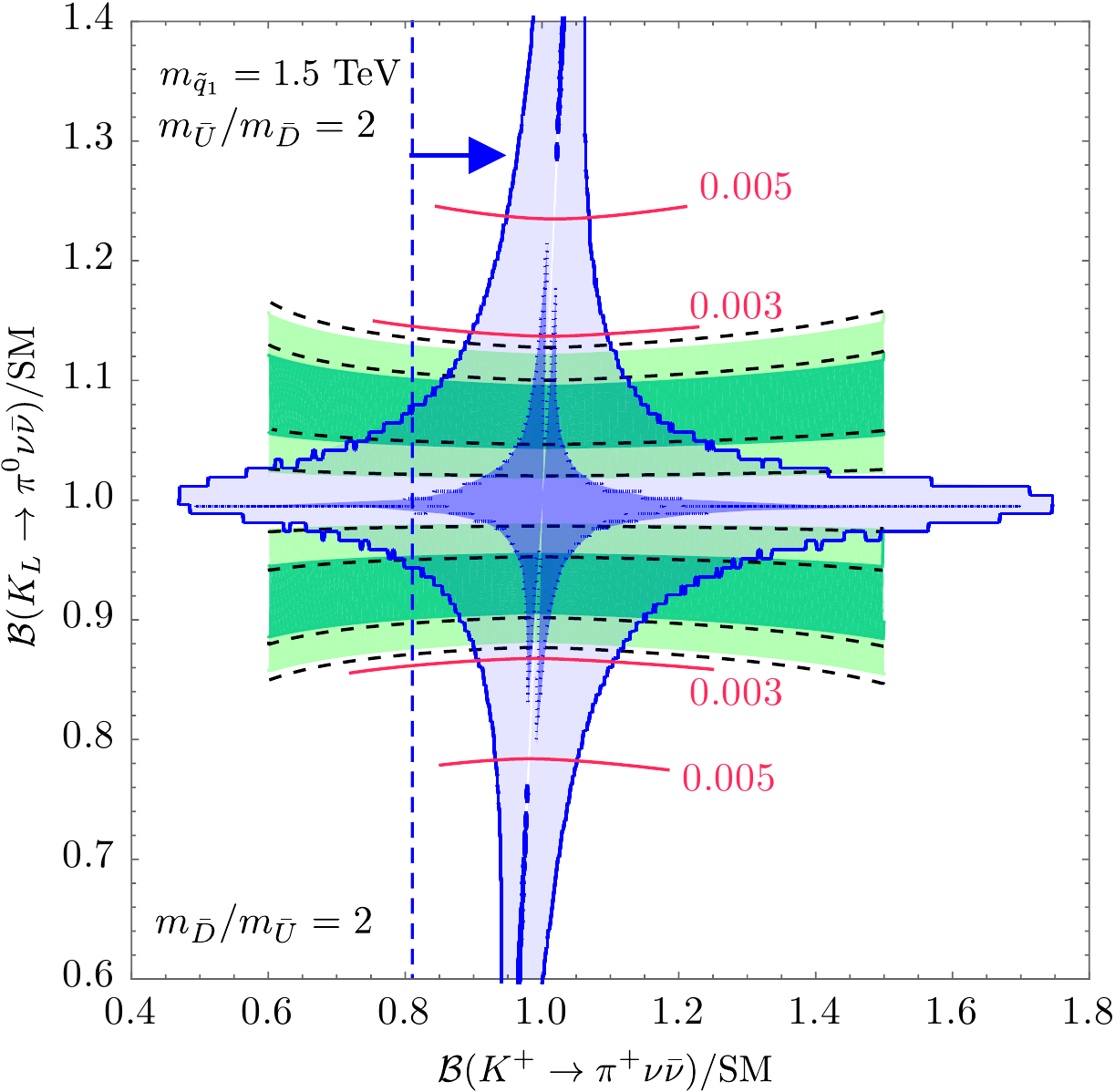}}
\caption[]{The correlation is shown in the Trojan penguin scenario.
The light (dark) blue region requires a milder parameter
  tuning than 1\,\%\,(10\,\%) of the gluino mass and  the CP violating phase in order to suppress contributions to $\epsilon_K$.  The red contour represents the SUSY contributions to
  $\epsilon^{\prime}_K / \epsilon_K$, and the
  $\epsilon_K^\prime/\epsilon_K$ discrepancy is resolved at
  $1\,\sigma$\,($2\,\sigma$) within the dark (light) green region. 
   The
  lightest squark mass is fixed to 1.5\,TeV.  In the \emph{left}
  panel, $m_{\bar{D}}/m_{\bar{U}} = 1.1$ ($m_{\bar{U}}/m_{\bar{D}} =
  1.1$) is used for $0 < \theta < \pi$ ($\pi < \theta < 2 \pi$) to
  obtain a positive SUSY contribution to $\epsilon^{\prime}_K /
  \epsilon_K$. While, $m_{\bar{D}}/m_{\bar{U}} = 2$
  ($m_{\bar{U}}/m_{\bar{D}} = 2$) is used for $0 < \theta < \pi$ ($\pi <
  \theta < 2 \pi$) in the \emph{right} panel.  The region on the right
  side of the blue dashed lines are allowed by the current experimental
  measurements. }
\label{fig:boxresult}
\end{figure}
%%%%%%%%%%%%%%%%%%%%%%%%%%%%%%%%%%%%%%%%%%%%

We first focus on the box dominated scenario, where all NP contributions to $|\Delta S|=1 $ and $ |\Delta S| = 2$ processes are dominated by the four-fermion box diagrams.
Such a situation is realized in the minimal supersymmetric standard model~\cite{Kitahara:2016otd}.
The desired effect in $\epsilon_K\primed$ is generated via gluino-squark box diagrams when the mass difference between the right-handed up and down squarks exists. 
Such a contribution is called {\it Trojan penguin} because its effect is parameterized by the electroweak penguin operator at low energy scale~\cite{gkn}.
While the sizable effects in $\epsilon_K\primed$ are obtained by the Trojan penguin,  a simultaneous efficient suppression of the supersymmetric QCD contributions to $\epsilon_K$ can be achieved.
The suppression occurs because crossed and uncrossed gluino box-diagrams cancel in $|\Delta S|=2 $  process, if the gluino mass is roughly $1.5$ times the squark masses. 
With appropriately large left-left squark mixing angle and a CP violating phase, one can reconcile the measurements of $\epsilon_K$,  $\Delta M_K$ and collider searches for the colored particles with the sizable contribution to $\epsilon_K\primed$.

However, there is no such cancellation in the (dominant) chargino box
contribution to $K_L\to \pi^0 \nu \overline{\nu}$ and $K^+\to \pi^+ \nu
\overline{\nu}$ which permits potentially large effects.
We investigate the correlation between $\epsilon_K\primed$ and $\mathcal{B}(K \to \pi \nu \overline{\nu})$ varying the following parameters:
\begin{equation}
|\Delta_{Q,12}|,~\theta,~M_3,~m_{\bar{U}}/m_{\bar{D}}, 
\end{equation}
 with $0 < |\Delta_{Q,12}| < 1$ and $0 < \theta < 2 \pi$. Here, defining the bilinear terms for the squarks as $M^2_{X,ij} = m_X^2
(\delta_{ij} + \Delta_{X,ij})$ for $X=Q,\bar U, \bar D$, $\theta \equiv \textrm{arg}(\Delta_{Q,12})$, $M_3$ is the
gluino mass. 
We fix the slepton mass and the lightest squark mass close to the experimental limit ($m_L
= 300\,$GeV and $m_{\tilde{q}_1} = 1.5\,$TeV) and use GUT
relations among all three gaugino masses.

The main result is shown in  Fig.~\ref{fig:boxresult}  in the $\mathcal{B}(K_L\to\pi^{0}
\nu\overline{\nu})$--$\mathcal{B}(K^{+}\to\pi^{+} \nu\overline{\nu})$
plane which are normalized by their SM predictions.
We find that  the necessary amount of the tuning in the
gluino mass and the CP violating phase   in order to suppress contributions to $\epsilon_K$  determines deviations of $\mathcal{B}(K \to\pi \nu\overline{\nu})$ from the SM values.
The current
$\epsilon_K^\prime/\epsilon_K$ discrepancy is resolved at
$1\,\sigma$\,($2\,\sigma$) within the dark (light) green region.  
In the left (right) panel we used
$m_{\bar{D}}/m_{\bar{U}} = 1.1$ (2) with $m_{\bar{U}} = m_Q$ for $0 <
\theta < \pi$, and $m_{\bar{U}}/m_{\bar{D}} = 1.1$ (2) with $m_{\bar{D}}
= m_Q$ for $\pi < \theta < 2 \pi$.  
Numerically, we observe $\mathcal{B}(K_L \to \pi^0\nu\overline{\nu})/\mathcal{B}^{\rm SM}
  (K_L \to \pi^0\nu\overline{\nu})\lesssim 2\,(1.2)$ and $\mathcal{B}(K^+ \to
  \pi^+\nu\overline{\nu})/\mathcal{B}^{\rm SM}(K^+ \to \pi^+\nu\overline{\nu})
  \lesssim 1.4\,(1.1)$ in light of $\epsilon^{\prime}_K / \epsilon_K$ discrepancy, 
  if all squarks are heavier than $1.5\,\tev$ and if a $1\,(10)\,\%$ fine-tuning is permitted.

We also observe a strict correlation between $\mathcal{B}(K_L \to
\pi^0\nu\overline{\nu})$ and $m_{\bar{U}}/m_{\bar{D}}$: $\mbox{sgn}\,
(\mathcal{B}(K_L \to \pi^0\nu\overline{\nu})-\mathcal{B}^{\rm SM} (K_L
\to \pi^0\nu\overline{\nu}) ) = \mbox{sgn}\,(m_{\bar{U}}-m_{\bar{D}}) $.
Thus, $\mathcal{B}(K_L \to \pi^0\nu\overline{\nu})$ 
can indirectly determine  whether  the right-handed up or  down squark is the heavier one.

\section{$Z$-penguin dominated (modified $Z$-coupling) scenario}

%%%%%%%%%%%%%%%%%%%%%%%%%%%%%%%%%%%%%%%%%%%%
\begin{figure}
\centerline{\includegraphics[width=0.4\textwidth, bb = 0 0 561 472]{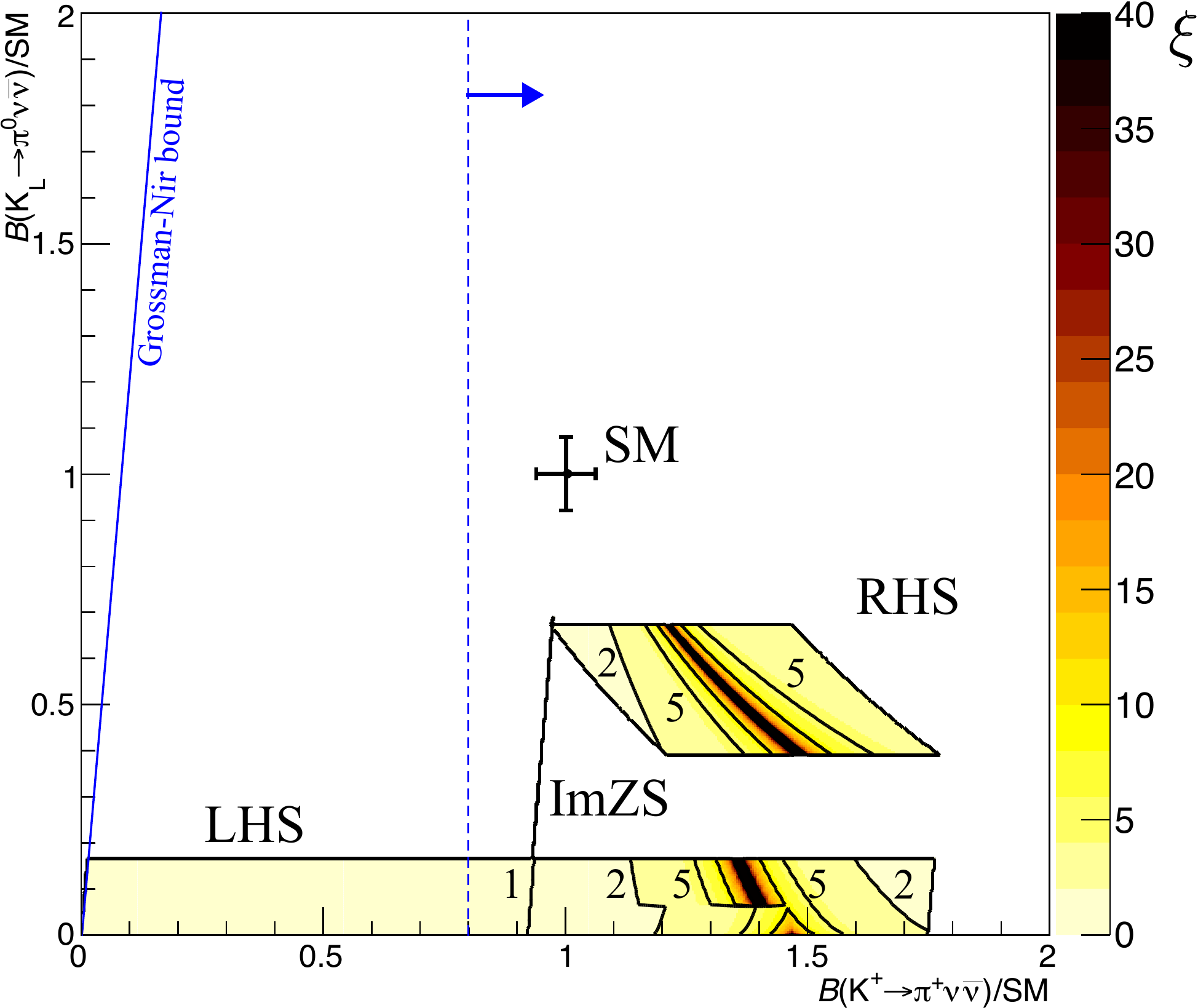}~~~~
\includegraphics[width=0.4\textwidth, bb = 0 0 558 472]{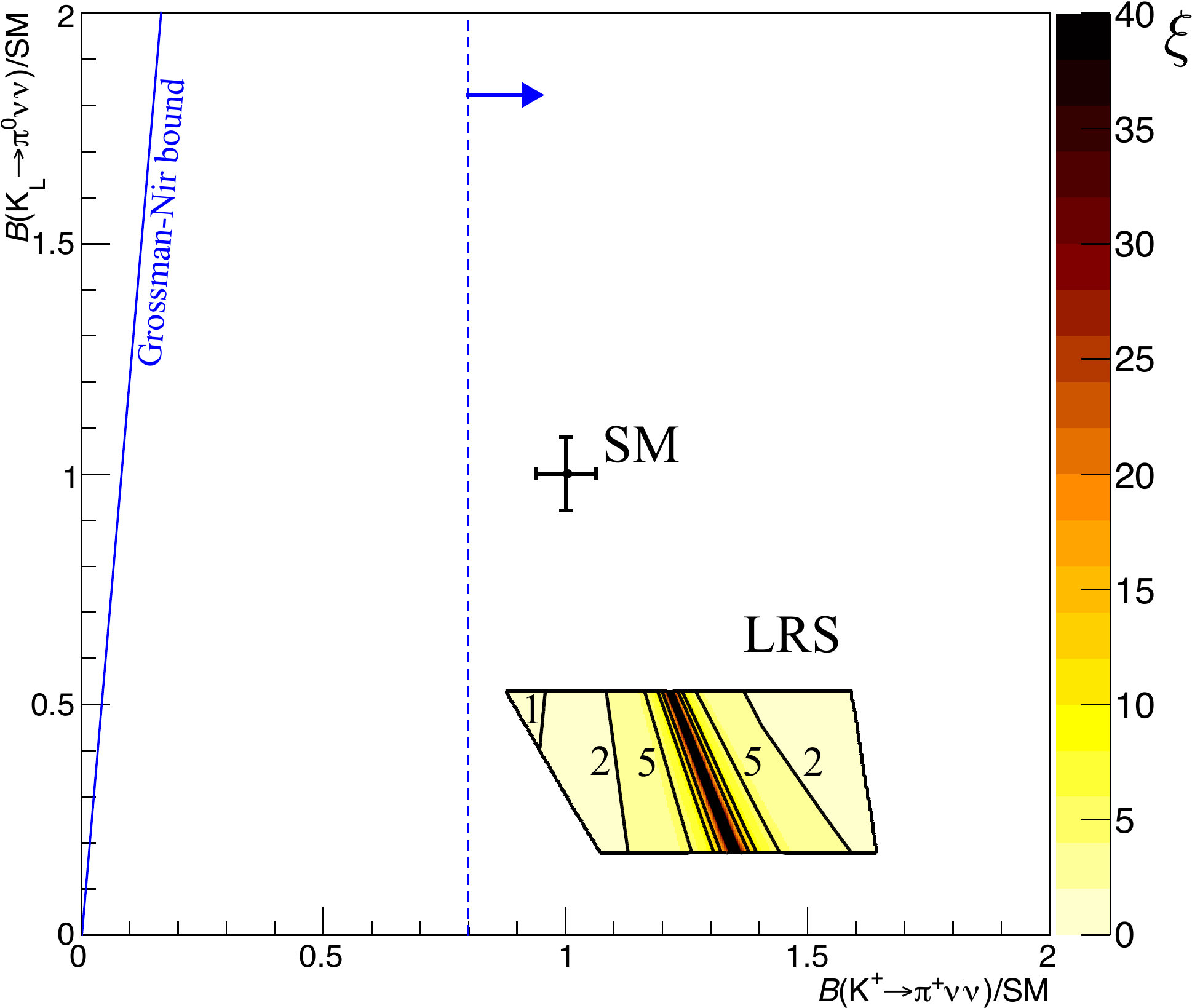}}
\caption[]{
Contours of the tuning parameter $\xi$ are shown in the simplified modified $Z$-coupling scenarios: LHS, RHS, and ImZS ({\em left} panel) and LRS ({\em right}).
In the colored regions, $\epsilon^{\prime}_K/ \epsilon_K$ is explained at $1\sigma$, and the experimental bounds of $\epsilon_K$, $\Delta M_K$, and $\mathcal{B}(K_L \to \mu^{+} \mu^{-})$ are satisfied.
The right region of the blue dashed line is allowed by the measurement of $\mathcal{B}(K^{+} \to \pi^{+} \nu \overline{\nu})$ at $1\sigma$.
The NP scale is set to be $\mu=1$\,TeV.
}
\label{fig:simplified}
\end{figure}
%%%%%%%%%%%%%%%%%%%%%%%%%%%%%%%%%%%%%%%%%%%%

Next, we focus on the $Z$-penguin dominated scenario. 
The negative dominant contribution to $\epsilon^{\prime}_K / \epsilon_K$ comes from 
 $Z$-penguin diagrams in the SM.
 Since in the SM there is a large numerical cancelation between QCD-penguin and the $Z$-penguin contributions to $\epsilon^{\prime}_K / \epsilon_K$, 
 a modified $Z$ flavor-changing ($s$--$d$) interaction from NP can explain the current $\epsilon^{\prime}_K / \epsilon_K$ easily \cite{Buras:2014sba}.
Then,  
the decay, $s\to d\nu  \overline{\nu}$, proceeding   through an intermediate $Z$ boson, is modified by  NP.
Therefore, the branching ratios of $K \to \pi\nu\bar\nu$ are likely to deviate from the SM predictions once the $\epsilon^{\prime}_K/\epsilon_K$ discrepancy is explained by the modified $Z$-coupling. 
They could be a signal to test the scenario.

Such a signal is constrained by $\epsilon_K$.
The modified $Z$ couplings affect $\epsilon_K$  via the so-called double penguin diagrams; the $Z$ boson mediates the transition with two  flavor-changing $Z$ couplings.
Such a contribution is enhanced when there are both left-handed  and right-handed couplings because of the chiral enhancement of the hadronic matrix elements. 
An important point is that 
since the left-handed coupling is already present in the SM, 
the right-handed coupling must be constrained even without NP contributions to the left-handed one. 
Such interference contributions between the NP and the SM have been overlooked in the literature. 
We~\cite{Endo:2016tnu} and recent work by Bobeth {\it et al}\,~\cite{Bobeth:2017xry} have revisited the modified $Z$-coupling scenario including the interference contributions, and found 
the parameter regions allowed by the indirect CP violation change significantly.

We find that similar to the previous section, the deviations of $\mathcal{B}(K \to \pi \nu \overline{\nu})$ from the SM values are  determined  by the necessary amount of the tuning in NP contributions  to $\epsilon_K$.
We parametrize it by $\xi$:  A degree of the NP parameter tuning is represented  by $1 / \xi$, 
e.g., $\xi = 10$ means that the model parameters are tuned at the 10\% level.

In Fig.~\ref{fig:simplified}, contours of the tuning parameter $\xi$ are shown for the simplified scenarios:  LHS (all NP effects  appear as left-handed), RHS (all NP effects  appear as right-handed), ImZS (NP effects are purely imaginary), and LRS (left-right symmetric scenario) on the plane of the branching ratios of $K \to \pi \nu \overline{\nu}$ which are normalized by their SM predictions.
We scanned the whole parameter space of the modified $Z$-coupling in each scenario, and selected the parameters where $\epsilon^{\prime}_K / \epsilon_K$ is explained at the $1\sigma$ level.
The experimental bounds from $\epsilon_K$, $\Delta M_K$, and $\mathcal{B}(K_L \to \mu^{+} \mu^{-})$ are satisfied. 
In most of the allowed parameter regions, $\xi = \mathcal{O}(1)$ is obtained.
Thus, one does not require tight tunings in these simplified scenarios. 
In the figures, $\mathcal{B}(K_{L}\to \pi^{0} \nu \overline{\nu} )$ is smaller than the SM value by more than 30\%. 
On the other hand, $\mathcal{B}(K^{+}\to \pi^{+}\nu \overline{\nu})$ depends on the scenarios. 
In LHS, we obtain $0 < \mathcal{B}(K^{+}\to \pi^{+}\nu \overline{\nu})/ \mathcal{B}(K^{+}\to \pi^{+}\nu \overline{\nu})_{\rm SM} < 1.8$. 
In RHS, $\mathcal{B}(K^{+}\to \pi^{+}\nu \overline{\nu})$ is comparable to or larger than the SM value, but cannot be twice as large.
In ImZS, the branching ratios are perfectly correlated and $\mathcal{B}(K^{+}\to \pi^{+}\nu \overline{\nu})$ does not deviate from the SM one.  
In LRS, $\mathcal{B}(K_{L}\to \pi^{0} \nu \overline{\nu} )$ does not exceed about a half of the SM value.
The more general situation is discussed in Ref.~\cite{Endo:2016tnu}.

\section{Conclusions}

We have presented the correlations between $\epsilon^{\prime}_K / \epsilon_K$, $\mathcal{B}(K_L\to \pi^0 \nu \overline{\nu})$, and $\mathcal{B}(K^+\to \pi^+ \nu \overline{\nu})$   
in the box dominated scenario and the $Z$-penguin dominated one.
 It is shown that  
the constraint from $\epsilon_K$  produces different correlations between two NP scenarios.
In the future,  measurements of $\mathcal{B}(K\to \pi \nu \overline{\nu})$ will be significantly improved. 
The NA62 experiment at CERN measuring $\mathcal{B}(K^+\to \pi^+ \nu \overline{\nu})$ is aiming to reach a precision of 10\,\% compared to the SM value already in 2018.  In order to achieve $5\%$ accuracy more time is needed. 
Concerning $K_L \to \pi^0 \nu \overline{\nu}$, the KOTO experiment at J-PARC  aims in a first step at  measuring  $\mathcal{B}(K_L \to \pi^0 \nu \overline{\nu})$ around the SM sensitivity.
Furthermore, the KOTO-step2 experiment will aim at 100 events for the SM branching ratio, implying
a precision of 10\,\% of this measurement.
Therefore, 
we conclude that when the $\epsilon^{\prime}_K / \epsilon_K$ discrepancy is explained by the NP contribution, NA62 experiment could probe whether a modified $Z$-coupling scenario is realized or not, and KOTO-step2 experiment can distinguish  the box dominated scenario and the simplified modified $Z$-coupling scenario.

\section*{Acknowledgments}

I would like to thank  
Giancarlo D'Ambrosio,  Andreas Crivellin,   Motoi Endo, Satoshi Mishima, Ulrich Nierste, Paul Tremper, and Kei Yamamoto  for fruitful collaborations on  the presented work. I also want to warmly thank the organizers of Moriond EW 2017 for giving me the opportunity to present these results  in a great conference.

%\section*{Appendix}
%
%We can insert an appendix here and place equations so that they are
%given numbers such as Eq.~\ref{eq:app}.
%\be
%x = y.
%\label{eq:app}
%\ee


\section*{References}

\begin{thebibliography}{99}

\bibitem{dual}
A.~J.~Buras and J.~M.~G\'erard,
  %``1/n Expansion for Kaons,''
\Journal{\NPB}{264}{371}{1986};
  %%CITATION = doi:10.1016/0550-3213(86)90489-X;%%
A.~J.~Buras  {\it et al},
  %``Large $N$ Approach to Kaon Decays and Mixing 28 Years Later: $\Delta I = 1/2$ Rule, $\hat B_K$ and $\Delta M_K$,''
 {\it Eur.\ Phys.}\ J.\ C  {\bf 74} 2871 (2014) 
  [arXiv:1401.1385 [hep-ph]].
  %%CITATION = doi:10.1140/epjc/s10052-014-2871-x;%%
  
  \bibitem{Bai:2015nea} 
  Z.~Bai {\it et al} [RBC and UKQCD Collaborations],
  %``Standard Model Prediction for Direct CP Violation in $K \to \pi \pi$ Decay,''
 \Journal{\PRL}{115}{212001}{2015} 
  [arXiv:1505.07863 [hep-lat]].
  %%CITATION = doi:10.1103/PhysRevLett.115.212001;%%

  
    \bibitem{Kitahara:2016nld} 
  T.~Kitahara {\it et al},
  %``Singularity-free next-to-leading order $\Delta$S = 1 renormalization group evolution and $\epsilon_K'/\epsilon_K$ in the Standard Model and beyond,''
 {\it J. High Energy Phys.} {\bf 1612}, 078 (2016)
  [arXiv:1607.06727 [hep-ph]].
  
%\cite{Buras:2015yba}
\bibitem{Buras:2015yba}
A.~J. Buras {\it et al},
  %``Improved anatomy of ε′/ε in the Standard Model,''
{\it J. High Energy Phys.} {\bf 1511}, 202  (2015)
  %doi:10.1007/JHEP11(2015)202
  [arXiv:1507.06345 [hep-ph]].
  %%CITATION = doi:10.1007/JHEP11(2015)202;%%

\bibitem{Kitahara:2016ftn} 
  T.~Kitahara  {\it et al},
  %``Recent progress on CP violation in $K\to \pi \pi$ decays in the SM and a supersymmetric solution,''
{\it J.\ Phys.\ Conf.\ Ser.}\,{\bf 800}, 012019 (2017)
  [arXiv:1611.08278 [hep-ph]].
  
  
\bibitem{Olive:2016xmw} 
  C.~Patrignani {\it et al} [Particle Data Group],
  %``Review of Particle Physics,''
{ Chin.\ Phys.\ C {\bf 40},  100001 (2016).}
  % doi:10.1088/1674-1137/40/10/100001
  %%CITATION = doi:10.1088/1674-1137/40/10/100001;%%
  %281 citations counted in INSPIRE as of 18 Jan 2017
  
\bibitem{Lehner:2015jga}
  C.~Lehner  {\it et al},
  %``Emerging lattice approach to the K-Unitarity Triangle,''
  Phys.\ Lett.\ B {\bf 759} 82  (2016) 
  %doi:10.1016/j.physletb.2016.04.064
  [arXiv:1508.01801 [hep-ph]].
  %%CITATION = doi:10.1016/j.physletb.2016.04.064;%%
  
      
    \bibitem{Kitahara:2016otd} 
  T.~Kitahara {\it et al},
  %``Supersymmetric Explanation of CP Violation in $K\to \pi\pi$ Decays,''
  \Journal{\PRL}{117}{091802}{2016}
  [arXiv:1604.07400 [hep-ph]].
  
  \bibitem{gkn}
%\bibitem{Kagan:1999iq}
  A.~L.~Kagan and M.~Neubert,
  %``Large Delta I = 3/2 contribution to epsilon-prime / epsilon in supersymmetry,''
\Journal{\PRL}{83}{4929}{1999} 
  %doi:10.1103/PhysRevLett.83.4929
  [hep-ph/9908404].
  
\bibitem{Crivellin:2017gks} 
  A.~Crivellin {\it et al},
  %``$K\to \pi \nu\overline{\nu}$ in the MSSM in the Light of the $\epsilon^{\prime}_K/\epsilon_K$ Anomaly,''
  arXiv:1703.05786 [hep-ph].
  
      %\cite{Buras:2014sba}
\bibitem{Buras:2014sba} 
  A.~J.~Buras {\it et al},
%  ``$\Delta I=1/2$ rule, $\epsilon '/\epsilon $ and $K\rightarrow \pi \nu \bar{\nu }$ in $Z' (Z)$ and $G' $ models with FCNC quark couplings,''
{\it Eur.\ Phys.}\ J.\ C {\bf 74}, 2950 (2014)
  [arXiv:1404.3824 [hep-ph]];
  %%CITATION = doi:10.1140/epjc/s10052-014-2950-z;%%
  %\cite{Buras:2015yca}
%\bibitem{Buras:2015yca} 
  A.~J.~Buras {\it et al},
 % ``$ K\to \pi \nu \overline{\nu} $ and $\epsilon'/\epsilon$ in simplified new physics models,''
 {\it J. High Energy Phys.}  {\bf 1511}, 166 (2015)
  [arXiv:1507.08672 [hep-ph]];
  %%CITATION = doi:10.1007/JHEP11(2015)166;%%
  %
  %\cite{Buras:2015jaq}
%\bibitem{Buras:2015jaq} 
  A.~J.~Buras,
 % ``New physics patterns in $\epsilon^\prime/\epsilon$ and $\epsilon_K$ with implications for rare kaon decays and $\Delta M_K$,''
 {\it J. High Energy Phys.} {\bf 1604}, 071 (2016)
  [arXiv:1601.00005 [hep-ph]].
  %%CITATION = doi:10.1007/JHEP04(2016)071;%%
  
  
  \bibitem{Endo:2016tnu} 
  M.~Endo {\it et al},
  %``Revisiting Kaon Physics in General $Z$ Scenario,''
  arXiv:1612.08839 [hep-ph], to appear in {\it Phys. Lett.} B.
  
  %\cite{Bobeth:2017xry}
\bibitem{Bobeth:2017xry} 
  C.~Bobeth {\it et al},
  %``Yukawa enhancement of $Z$-mediated New Physics in $\Delta S = 2$ and $\Delta B = 2$ Processes,''
  arXiv:1703.04753 [hep-ph].
  %%CITATION = ARXIV:1703.04753;%%
  %4 citations counted in INSPIRE as of 14 May 2017

  

  

\end{thebibliography}
\end{document}